\documentclass[conference]{IEEEtran}
\IEEEoverridecommandlockouts

\usepackage{nicefrac}
\usepackage[left=2.5cm,right=2.5cm,top=1.7cm]{geometry}
\usepackage[none]{hyphenat}

\usepackage{cite}
\usepackage{amsmath,amssymb,amsfonts}
\usepackage{algorithmic}
\usepackage{graphicx}
\usepackage{textcomp}
\usepackage{xcolor}
\def\BibTeX{{\rm B\kern-.05em{\sc i\kern-.025em b}\kern-.08em
    T\kern-.1667em\lower.7ex\hbox{E}\kern-.125emX}}

\begin{document}

\title{Open RAN Slicing with Quantum Optimization\\
}

\author{\IEEEauthorblockN{Patatchona Keyela, Soumaya Cherkaoui}
\IEEEauthorblockA{\textit{Computer Engineering and Software Engineering Dept.} \\
\textit{Polytechnique Montreal, Montreal, Canada}\\
\{keyela.patatchona, soumaya.cherkaoui\}@polymtl.ca}
}

\maketitle

\begin{abstract}
RAN slicing  technology is a key aspect of the  Open RAN paradigm, allowing simultaneous and independent provision of various services such as ultra-reliable low-latency communications (URLLC), enhanced mobile broadband (eMBB), and massive machine-type communications (mMTC) through virtual networks that share a single radio access infrastructure. Efficient resource allocation is crucial for RAN slicing, as each service has specific quality of service (QoS) requirements, and a balance between different services must be maintained. Although  heuristic and reinforcement learning (RL) techniques  have been explored to achieve efficient resource allocation, these approaches face notable limitations: heuristic algorithms face complexity issues that limit their effectiveness in large networks, RL solutions are constrained by their dependency on training data and struggle to adapt to new scenarios and environments. This paper proposes a framework that leverages quantum optimization techniques to optimize radio resource blocks allocation in Open RAN slicing for URLLC and eMBB services. We provide a classical problem formulation and the quantum implementation using the constrained quadratic model on Dwave quantum annealing platform, showcasing the potential of quantum optimization techniques to deliver in real-time optimal solutions for optimization problems in 5G and beyond networks.
\end{abstract}


\section{Introduction}
\label{sec:intro}

The 5G and next-generation wireless networks aim to enable a range of applications, from autonomous vehicles, virtual reality to remote surgery, each requiring specific service quality. To meet these demands, network architectures must be more flexible, scalable, and capable of meeting various quality-of-service (QoS) requirements. Open Radio Access Network (Open RAN) is a relatively new paradigm within modern telecommunications that aims to meet these needs by offering modular networks, vendor-agnostic, with support of network slicing suitable for different service types\cite{electronics12102200}.

The Open RAN architecture, though designed to be flexible and programmable, is being challenged by effective resource allocation between different service slices, each having strict QoS requirements on latency and data rates. Balancing these requirements in real-time across a shared infrastructure is computationally complex, especially as the size and heterogeneity of networks increase. Traditional approaches to this problem include heuristic algorithms, such as dynamic and linear programming, which often lack scalability and adaptability. Deep learning-based solutions have also been applied to resource allocation problems, but they are highly dependent on training data, which limits their adaptability to new network conditions.

Among significant research that has been conducted on radio resource allocation to improve flexibility in RAN slicing, \cite{9888767} addresses service-aware allocation in Open RAN, targeting limited fronthaul capacity and strict delay constraints by iteratively allocating power, assigning PRBs, and selecting virtual network functions (VNF), all followed by a greedy  Open RAN radio unit (O-RU) association method. This improves data rates and reduces delays but the scalability remains an issue.
In \cite{8958689}, RAN slicing architecture for URLLC uses multi-access edge computing and a platform as a service (PaaS) model for dynamic resource management to achieve low-latency, high-reliability service but struggles with real-time operations and high user mobility.
Similarly, \cite{3480842} introduces NexRAN framework, which applies a closed-loop RAN slicing strategy within the Open RAN ecosystem to support real-time resource allocation adjustments via an xApp on the RAN Intelligent Controller (RIC), yet it lacks robustness and scalability for broader applications.
Techniques like priority-based resource allocation \cite{8418083} maximize Quality of Experience (QoE) under constraints of limited bandwidth and data rate. Genetic algorithms have also been applied in scenarios requiring energy efficiency in downlink orthogonal frequency-division multiple access (OFDMA) heterogeneous networks \cite{9631283}, but they require substantial computation and parameter tuning to manage the complexity of mixed-integer nonlinear programming problems.

Deep RL (DRL) offers a learning-based alternative for resource allocation policies: \cite{9771605} proposes two DRL models that balance energy consumption and latency to optimize resource allocation in a dynamically sliced Open RAN environment, outperforming greedy baselines but they demande significant training data. Federated DRL \cite{9999295} enhances scalability and privacy by aggregating at the RIC locally trained models by mobile virtual network operators, improving QoS for eMBB and URLLC users. Nevertheless, DRL approaches still rely on extensive data and high computational overhead, underscoring the need for more efficient methods.

In this study, we propose a quantum annealing (QA) framework for Open RAN resource allocation, formulated as a constrained quadratic model (CQM) for physical resource blocs (PRBs) allocation in a multi-slice Open RAN configuration, capable of achieving optimal resource allocation within a reasonable timeframe for real-time applications. Through this framework, we aim not only to demonstrate that quantum optimization is able to meet the diverse QoS demands of different service slices and provide (near) optimal solutions quickly enough for real-time applications, but we also analyze its scalability  for next-generation wireless networks. Our contributions are as follows:
\begin{itemize}
    \item We formulate the resource allocation problem specifically for Open RAN architecture, accounting for its unique constraints and requirements.
    \item We present a quantum-native formulation of this problem and implement it on a D-Wave quantum computer, showcasing the feasibility of using quantum computing for real-time network optimization.
    \item We conducted a realistic simulation of slicing in Open RAN  with a detailed analysis of the strengths and limitations of quantum annealing as an optimization technique in the context of network resource allocation problems.
\end{itemize}

This rest of the paper is organized as follows: section \ref{sec:system-model} presents the system model, section \ref{sec:prob-formulation} describes the classical problem formulation and the CQM for resource allocation, section \ref{sec:results} details the simulation setup, the quantum annealing implementation on the D-Wave platform and discusses the results. Finally, section \ref{sec:conclusion} summarizes key findings, contributions, and future research directions.

\section{System Model}
\label{sec:system-model}

\begin{figure}[htbp]
    \centering
    \includegraphics[width=0.48\textwidth]{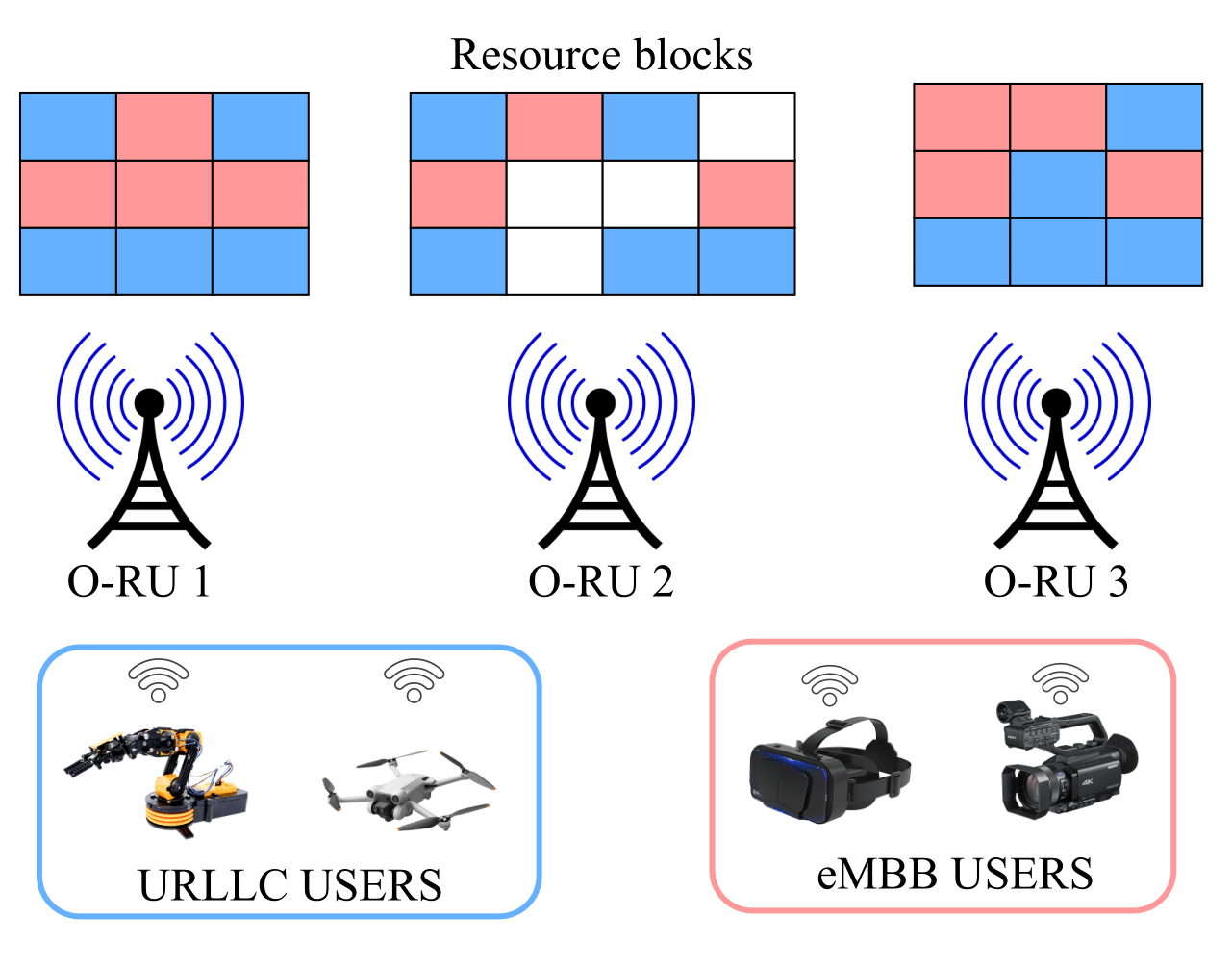}
    \caption{System model for resource blocks}
    \label{fig:system-model}
\end{figure}

We consider the problem of physical resource blocks allocation  in an Open RAN  system delivering latency sensitive services and services requiring high data rates through network slicing technology, similar to the model described in \cite{9729992}. We partition the RAN into URLLC slice and eMBB slice and users \(\mathcal{U} = \{1, 2, 3, \dots, N\}\) are being served by a set of \(M\) gNodeBs \(\mathcal{M} = \{1, 2, \ldots, M\} \) as illustrated in Fig.\ref{fig:system-model}. All the users are uniformly distributed in the area of the network coverage and every user requests only one service at a time and is served through only one gNodeB. 

In this model, we consider \( \mathcal{U}_m\) to be the set of users already associated with the gNodeB \(m\). For each service request, the scheduler allocates a maximum of \(K_{max}\) available resource blocks (RBs) to the user. If no RB is available, the gNodeB borrows from another gNodeB within the network to meet the user’s QoS requirements in terms of data rate and latency. Each gNodeB \(m\) has \(K_m\) RBs, but for simplicity, we consider the RBs of all gNodeBs as the set \( \mathcal{K} = \{1, 2, \ldots, K_1, K_1+1, \ldots, K\}\).

We introduced binary variables \( x_{kmn}\), representing the assignment of an RB \( k \) from gNodeB \(m\) to end-user \( n \), with values of \(1\) for allocation and \(0\) otherwise. Since a RB can only be assigned to one end-user, an additional RB request from gNodeB \(m\) to another gNodeB \(m'\) requires that the borrowed RB is idle. To ensure that each RB \( k \) is assigned to only one end-user \( n \) across the entire network, the following constraint has to be enforced:
\begin{equation}
    \sum_{n \in \mathcal{U}} x_{kmn} \leq 1, \quad \forall k \in \mathcal{K}.
\end{equation}

Given the bandwidth \( W \) of an RB,  the transmission power \( P_{smn} \) to end-user \(n\) of gNodeB \(m\) requesting service type \(s\) (\(s\) represents either the URLLC  or the eMBB slice),  the channel gain \( G_{kmn} \), and  the additive white Gaussian noise power \( P_{noise} \), the data rate achieved by end-user \( n \) when served by RB \( k\) is given by (\ref{eq:rkn}) \cite{9729992}:
\begin{equation}
\label{eq:rkn}
    r_{kmn} = W \log_2 \left( 1 + \frac{P_{smn} G_{kmn}}{P_{noise}} \right).
\end{equation}

Since the end-user \(n\) of gNodeB \(m\) can be served by multiple RBs simultaneously, the total data rate achieved is calculated by summing the data rates of all RBs serving the user:

\begin{equation}
\label{eq:rmn}
    r_{mn} = \sum_{k \in \mathcal{K}} x_{kmn} r_{kmn}.
\end{equation}

Note that formula (\ref{eq:rmn}) accounts for the possibility of borrowing RBs from other gNodeBs in the network. However, before any borrowing occurs, all RBs from the user's gNodeB must be exhausted, this condition can be enforced as follows:
\begin{equation}
\label{eq:xkprmn}
    x_{k'mn} |\mathcal{K}_m| \leq \sum_{n \in \mathcal{U}_m} \sum_{k \in \mathcal{K}_m} x_{kmn}, \quad \forall k' \in \mathcal{K} \setminus \mathcal{K}_m.
\end{equation}

The packet delay experienced by an end-user \( n \) depends on the traffic type, and we estimate it under the following assumptions: first, data packets from the gNodeB to the user follow a Poisson distribution, and second, the duration between the arrival of two consecutive packets is independent and exponentially distributed. Applying Little’s Law (relationship between the average number of items \(L\), the average arrival rate \(\lambda\) and the average service time \(\Delta t\)) \cite{Serfozo1999} to our model, which can be considered an \(M|M|1\) system, the packet delay for user \( n \) is estimated as follows:
\begin{equation}
    d_{mn} = (\delta_s^{-1} r_{mn} - \lambda_{s})^{-1},
\end{equation}
\noindent where \( \lambda_{s} \) is the packet arrival rate for the user’s slice, \( r_{mn} \) is the data rate, and \(\delta_s\) is the packet length in bits.

\section{Problem Formulation}
\label{sec:prob-formulation}
\subsection{Classical Problem Formulation}
When formulating the problem of RB allocation in Open RAN slicing, we consider not only the constraints described above, but also QoS requirements for both URLLC and eMBB slices in terms of data rate and packet delay.
\begin{gather}
\label{eq:cl_obj}
\mathcal{O}bj: \quad \max_{x} \sum_{m \in \mathcal{M}} \sum_{n \in \mathcal{U}_m} r_{mn}
\end{gather}
subject to:
\begin{align*}
&\mathcal{C}_1: \quad \sum_{k \in \mathcal{K}} x_{kmn} \leq K_{\max}, \quad \forall m \in \mathcal{M}, \forall n \in \mathcal{U}_m, \\
&\mathcal{C}_2:\quad\sum_{m \in \mathcal{M}} \sum_{n \in \mathcal{U}_m} x_{kmn} \leq 1, \quad \forall k \in \mathcal{K}, \\
&\mathcal{C}_3:\quad  x_{k'mn} |\mathcal{K}_m| \leq \sum_{n \in \mathcal{U}_m} \sum_{k \in \mathcal{K}_m} x_{kmn},\\
&\forall k' \in \mathcal{K} \setminus \mathcal{K}_m, \forall n \in \mathcal{U}_m, \forall m \in \mathcal{M}, \\
&\mathcal{C}_4:\quad r_{mn} \geq R_{\min}, \quad \forall n \in \mathcal{U}_m, \forall m \in \mathcal{M}, \\
&\mathcal{C}_5:\quad d_{mn} \leq D_{\max}, \quad \forall n \in \mathcal{U}_m, \forall m \in \mathcal{M}, \\
&\mathcal{C}_6:\quad x_{kmn} \in \{0, 1\}, \quad \forall k \in \mathcal{K}, \forall n \in \mathcal{U}_m, \forall m \in \mathcal{M}.
\end{align*}

The objective function (\ref{eq:cl_obj}) aims to maximize the overall data rate achieved by both URLLC and eMBB end-users. \(\mathcal{C}_1\) is introduced to ensure fairness in RB allocation by limiting the number of RBs allocated to each end-user to a maximum value \(K_{\max}\). \(\mathcal{C}_2\) ensures that each RB is assigned to at most one end-user at any given time. \(\mathcal{C}_3\) manages additional RB allocation requests, enforcing the constraint (\ref{eq:xkprmn}). \(\mathcal{C}_4\) and \(\mathcal{C}_5\) represent the QoS requirements for data rate and delay, respectively, ensuring that eMBB users receive a data rate above a minimum threshold \(R_{\min}\) and that URLLC users experience packet delays below a maximum threshold \(D_{\max}\). Finally, \(\mathcal{C}_6\) defines the allocation variables as binary.

The problem as formulated above is NP-hard and we prove it by relaxing the problem formulation and removing the constraints $\mathcal{C}_3$, $\mathcal{C}_4$, and $\mathcal{C}_5$. The simplified version becomes a $0/1$ knapsack problem:
\begin{align}
\label{eq:knapsack}
&\max_{x} \sum_{m \in \mathcal{M}} \sum_{n \in \mathcal{U}_m} r_{mn}\\
&\text{s.t.  }\sum_{k \in \mathcal{K}} x_{kmn} \leq K_{\max}, \quad \forall m \in \mathcal{M}, \forall n \in \mathcal{U}_m,\notag \\
&\sum_{m \in \mathcal{M}} \sum_{n \in \mathcal{U}_m} x_{kmn} \leq 1, \quad \forall k \in \mathcal{K},\notag \\
&x_{kmn} \in \{0, 1\}, \quad \forall k \in \mathcal{K}, \forall n \in \mathcal{U}_m, \forall m \in \mathcal{M}.\notag
\end{align}

In this reformulation, users $u_{mn}$ correspond to objects in the $0/1$ knapsack problem, the data rate $r_{mn}$ are their values, the cost for every user the number of allocated RBs, and the overall capacity is $|\mathcal{K}|$. This establishes the NP-hardness of (\ref{eq:knapsack}) and, by extension, of (\ref{eq:cl_obj}). To solve this problem we propose a QA-based approach, as quantum techniques hold the potential to find optimal or near-optimal solutions within time frames suitable for real-time dynamic resource allocation, independent of environmental conditions.

\subsection{Quantum Annealing}
QA is an optimization technique that uses principles of quantum mechanics to solve complex optimization problems. The fundamental mechanism of quantum annealing involves the use of quantum fluctuations to explore the solution space more efficiently than classical algorithms, which is particularly beneficial for NP-hard problems \cite{access2023/3271969, access2022/3188117}.

To solve an optimization problem, the quantum annealer initializes a quantum system in a superposition of states and then evolves it towards a low-energy state that corresponds to the optimal solution of the problem. The quantum system is described by a time-dependent Hamiltonian
\[
H(t) = \left(1 - \frac{t}{T}\right) H_{\text{init}} + \frac{t}{T} H_{\text{final}},
\]
which represents the energy of a particular state of the system. Here, \( H_{\text{init}} \) is the initial state where all qubits are in a superposition of \( |0\rangle \) and \( |1\rangle \), \( T \) is the total annealing time of the system, and \( t \in [0, T] \). This process is guided by an annealing schedule that dictates how the system transitions from a superposition to a definite state, significantly influencing the performance of the quantum annealer \cite{access2023/3271969, access2023/3318206}.

By leveraging quantum tunneling and thermal fluctuations, quantum annealers are highly effective for solving complex and large-scale optimization problems, making certain NP-hard problems more tractable \cite{11128576}. QA provides consistent results and reduces the variability typically observed in heuristic approaches \cite{Yarkoni_2022}, serving as an efficient technique to tackle resource allocation optimization challenges in Open RAN slicing.

\subsection{Quantum Optimization Models}
Developing a quantum optimization model to run on a quantum annealer requires first clearly defining the problem by identifying the objective and all constraints, making a mathematical formulation. The next step is to translate this formulated problem into a format compatible with a quantum annealer, which is the focus of this subsection.

We build a QA model for our problem to execute it  on D-Wave's cloud-based platform, ant D-Wave's QPU support various input models for optimization problems, including the Binary Quadratic Model (BQM), Constrained Quadratic Model (CQM), Discrete Quadratic Model (DQM), and the recently introduced Non-Linear Sampler \cite{dimod_models_reference}. These models differ primarily in the types of variables they support, and the term "quadratic" refers to the maximum degree of interaction between variables being two. 

Traditionally, such problems are translated into a Quadratic Unconstrained Binary Optimization (QUBO) model, where the variables are binary (\(0/1\)) and there are no constraints, only a single minimization objective function expressed as a sum of quadratic terms. This can be represented as \(\min_x x^TQx\), where \(x\) is a vector of binary variables, and \(Q\) is the interaction matrix. However, for our problem formulated in (\ref{eq:cl_obj}), we choose to use a CQM, which we send to the quantum annealer for sampling. 

Although the decision variables in (\ref{eq:cl_obj}) are all binary and our constraints are linear, the CQM solver has the advantage of natively supporting both equality and inequality constraints. This feature offers a significant benefit when working with constrained optimization problems in QA, as it removes the need to convert constraints into penalty terms, which would require adding slack variables and thereby increase the number of variables, particularly when dealing with inequalities. One of the main challenges with other models is the need to set penalty coefficients appropriately for the constraints, while CQM effectively eliminates this challenge.

\section{Simulation Results}
\label{sec:results}

\subsection{Network Simulation and Configuration}
In this section, we demonstrate the results of implementing and running the proposed algorithm on D-Wave’s cloud-based CQM hybrid solver. All the results presented here were obtained a model we built using a computer with the following specifications: Intel(R) Core(TM) i7-10750H CPU @ 2.60 GHz, Windows 11 Professional, and 16 GB of RAM. 
\begin{table}[htbp]
    \caption{Simulation Parameters} 
    \label{parameters} 
    \centering
    \begin{tabular}{|p{5.3cm}|p{1.6cm}|} 
        \hline
        \textbf{Parameter} & \textbf{Value} \\ \hline
        Number of gNodeBs & 2--5 \\ \hline
        Total number of end-users & 8--150 \\ \hline
        Bandwidth of an RB & 180 KHz \\ \hline
        Transmit power of gNodeB, \(P_{kmn}\) & 30 dBm \\ \hline
        Power of AWGN, \(\sigma^2\) & -117 dBm \\ \hline
        Packet arrival rate per end-user, \(\lambda_s\) & 100 packets/s \\ \hline
        Packet length for eMBB \& URLLC  & 400 \& 120bits \\ \hline
        Minimum data rate for eMBB end-user, \(R_{\min}\) & 100 kbps \\ \hline
        Maximum delay for URLLC end-user, \(D_{\max}\) & 10 ms \\ \hline
    \end{tabular}
\end{table}

For the simulation, we adopted the commonly used 5G frequency \(n77\) band, corresponding to \(3.7\) GHz. The Open RAN O-RUs are assumed to have a coverage radius of \(300\) meters to approximate real-world conditions, and are placed within a \(1 \, \text{km} \times 1 \, \text{km}\) area. The users served by an O-RU are uniformly distributed within its coverage area, with each user requesting  a service from a single slice at the time. For simplicity, we kept the same value of the transmit power across slices and used the free-space path loss (FSPL) model to compute the channel gain:
\begin{equation*}
G_{u_b k} = \left( \frac{\lambda}{4 \pi d_{u_b k}} \right)^2,
\end{equation*}
where \(d_{u_b k}\) is the distance between the gNodeB and the user, and \(\lambda\) is the signal wavelength, calculated as \(\lambda = \frac{c}{f}\), with \(c\) being the speed of light (\(3 \times 10^8\) meters/second) and \(f\) the signal frequency. The remaining parameters, presented in Table \ref{parameters}, are chosen to reflect real-world values \cite{9729992}.

The network simulation replicates a realistic multi-slice deployment of gNodeBs and includes the following configuration: the network comprises \( M \) gNodeBs, each with its O-RU positioned at specific coordinates. The scheduler allocates a unique set of RBs to every gNodeB, ensuring localized resource management. The user set in the simulation consists of multiple users, each characterized by their spatial coordinates, a service request indicating their required slice type, and an association with exactly one gNodeB. This setup facilitates an accurate representation of resource allocation, slice management, service delivery within the network, availability of resources, and diverse demands of the users.

\subsection{Results Analysis}
Below, we present for the illustration, a scenario with RBs allocation rendered by the quantum annealer, based on the CQM of the problem (\ref{eq:cl_obj}).

\begin{figure}[htbp]
    \centering
    \includegraphics[width=0.95\columnwidth]{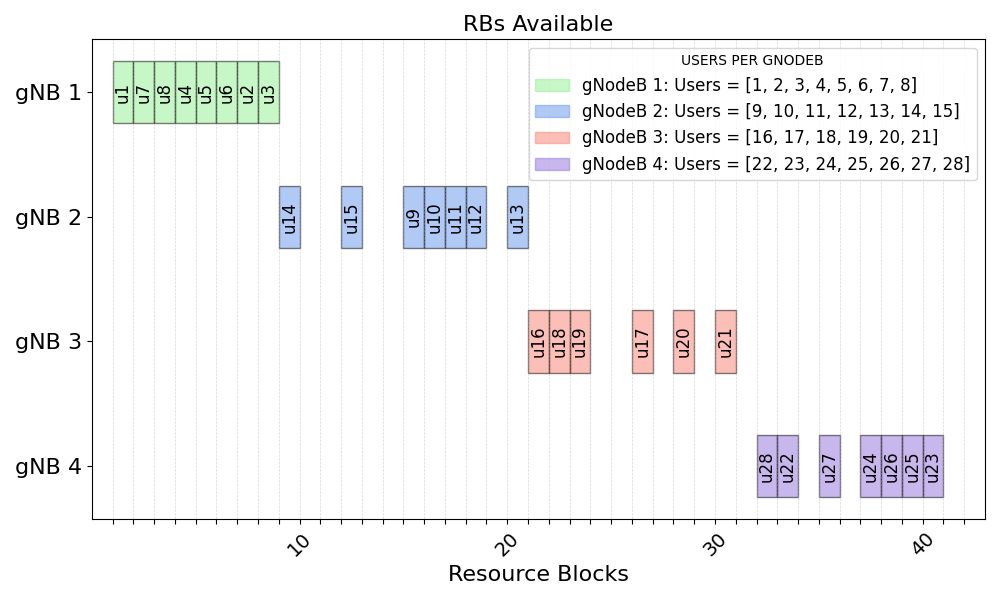}
    \caption{Scenario showing 4 gNBs, 42 RBs, and 28 users with resource blocks allocation by the quantum annealer}
    \label{fig:result1}
\end{figure}

The scenario in Fig.\ref{fig:result1} presents $4$ gNodeBs respectively having (RBs, users) \(\in \{(9 , 8), (12, 7), (11, 6), (10, 7)\}\). For each gNodeB, the number of RBs represents the total number of available RBs, and the number of users is the total service requests it receives from its primary users.
In the RBs allocation shown in Fig. \ref{fig:result1}, we observe that constraints \(\mathcal{C}_1\) and \(\mathcal{C}_2\) are successfully enforced: no user equipment exceeds the individual RB allocation limit, and no RB simultaneously serves two or more user equipment. In our model, RB borrowing constraint (\(\mathcal{C}_3\)), which maximizes the use of available resources, and all the other constraints of the problem are successfully enforced since we kept them as hard constraints during the model implementation. 

Important limitations have been observed during the simulation process, and the first is related to the quantum hardware capacity: the hybrid solver from D-Wave used to run the simulations is designed to optimally support up to $5000$ variables \cite{dwave_cqm_solver}, which roughly translates to handling scenarios of a network with up to $120$ RBs and $40$ users. 

For context, a single O-RU in an Open RAN setup, such as one provided by Benetel \cite{benetel_ran650}, operates with a bandwidth of $100$MHz. With a subcarrier spacing of $15$ kHz, such a radio can provide approximately $550$ physical RBs. In scenarios with $5$ gNodeBs, this scales to around $2750$ RBs to be managed, and for a number of users as low as $200$, the number of decision variables in the problem (\ref{eq:cl_obj}) significantly exceeds the current capacity of this quantum annealer, making it not very efficient.

To more objectively evaluate the performance limits of the solver  under stressed conditions, we intentionally violated the variables number limit by modeling scenarios extending it to $15000$ decision variables. Allocations were still feasible (and optimal or  near optimal), but the allocation time increased considerably, reaching $66.225$ seconds. When attempting to handle even larger instances, the computer's memory failed to prepare the model for submission to the quantum annealer, demonstrating practical constraints in system resources for such large-scale problems.

We recall that the parameter \(K_{\max}\) is the maximum number of RBs that any user equipment can receive, ensuring fairness in resource allocation. But during simulations, users with the highest achievable data rates tend to monopolize \( K_{\text{max}} \) RBs until no RBs remain. Consequently, the number of users receiving resources per gNodeB is approximately \(\lfloor \nicefrac{K_m}{K_{\max}}\rfloor + 1\). Any leftover RBs are reallocated to users associated with other gNodeBs, but in cases of insufficiency, the best-positioned users take \( K_{\text{max}} \), leaving others without resources. This phenomenon occurs from the choice of the objective function, which aims to maximize the total sum of data rates across users. While effective for throughput optimization, this approach inherently favors users in favorable positions. An alternative approach could focus on maximizing the number of users receiving service, potentially distributing resources more equitably. Such a modification in the objective function would address the disparity introduced by the current allocation strategy and provide a more balanced service distribution.

\begin{figure}[htbp]
    \centering
    \includegraphics[width=0.8\columnwidth]{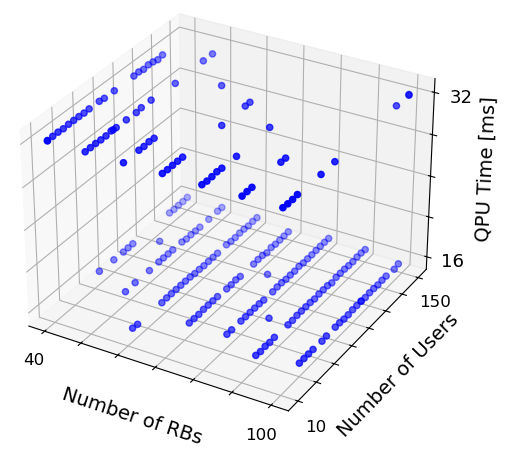}
    \caption{Annealing Time on the QPU}
    \label{fig:annealingTime}
\end{figure}


\begin{figure}[htbp]
    \centering
    \includegraphics[width=0.8\columnwidth]{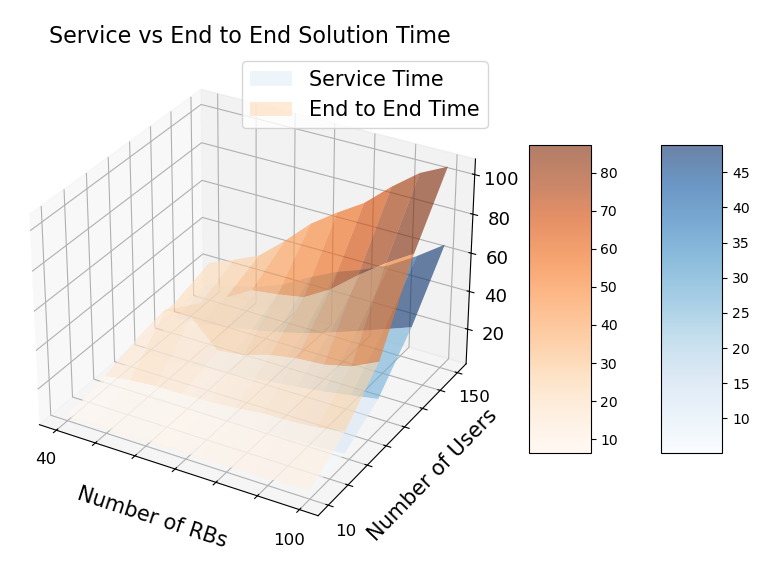}
    \caption{End-to-End Solution Time vs Service Time[sec.]}
    \vspace{-1.5 em}
    \label{fig:time-analysis}
\end{figure}

Fig. \ref{fig:annealingTime} illustrates the time it takes the QPU to execute quantum machine instructions of the submitted model, showing that the annealing process does not take more than $32$ ms. The service time on the hybrid solver, including queuing, preparation, sampling, and post-processing, is approximately $5$ seconds for small network configurations as illustrated in Fig.\ref{fig:time-analysis}. When the size of the network increases, the service progressively increases to reach $60$ seconds for a total of $100$ RBs and $150$ users, even though the annealing time remains $16$ or $32$ ms. This means the pre-processing and post-processing of the problem on the classic part of the solver are time-consuming, and as visible in Fig.\ref{fig:time-analysis}, the fact that the annealer is on the cloud adds latency due to the communication. Because we built the model with hard constraints, it is guaranteed that the packet delays for URLLC users and the data rates for eMBB users are within the required thresholds, however the allocation time fails the hypothesis of the application to real-time usecase mainly because of the harware limitations. This brings us to the conclusion that though QA has a great  potential for solving optimization problems, improvements on the hardware are still necessary for the real time applications. 

\section{Conclusion}
\label{sec:conclusion}

In this paper, we proposed an algorithm for resource block allocation in Open RAN systems, modeled it for QA, and demonstrated its implementation as a CQM on D-Wave’s cloud-based hybrid sampler. We conducted network simulations with various realistic scenarios of gNodeBs, RBs, and users, using parameters chosen to closely mimic real-world conditions. Our approach maximizes the total system throughput while meeting the stringent requirements for both eMBB and URLLC service types under realistic conditions. The results indicate that QA has the potential to effectively  address complex optimization problems inherent in resource allocation for 5G networks, offering optimal solutions within the different QoS requirements. The next step in this research work will be to build a deep learning model and a heuristic algorithm to have a comparative evaluation with quantum optimization.

\bibliographystyle{IEEEtran}
\bibliography{references}

\end{document}